\documentclass[british]{article}
\usepackage{ae,aecompl}
\usepackage[T1]{fontenc}
\usepackage[latin9]{inputenc}
\usepackage{geometry}
\usepackage{array}
\usepackage{multirow}
\usepackage{amsmath}
\usepackage{amsthm}
\usepackage{amssymb}
\usepackage{graphicx}
\usepackage{tablefootnote}
\usepackage{wasysym}

\makeatletter

\providecommand{\tabularnewline}{\\}

\date{}
\usepackage{array}

\makeatother

\usepackage{babel}
\begin{document}
\global\long\def\IN{\mathbb{N}}%
\global\long\def\II{\mathbbm{1}}%
\global\long\def\IZ{\mathbb{Z}}%
\global\long\def\IQ{\mathbb{Q}}%
\global\long\def\IR{\mathbb{R}}%
\global\long\def\IC{\mathbb{C}}%
\global\long\def\IP{\mathbb{P}}%
\global\long\def\IE{\mathbb{E}}%
\global\long\def\IV{\mathbb{V}}%

\title{Deviations from the Majority: A Local Flip Model}
\author{Gabor Toth\thanks{FernUniversit\"at in Hagen, Germany, gabor.toth@fernuni-hagen.de}
\;and Serge Galam\thanks{CEVIPOF - Centre for Political Research, Sciences Po and CNRS, Paris,
France, serge.galam@sciencespo.fr}}
\maketitle
\begin{abstract}
We study the effect of probabilistic distortions to the local majority
rules used in the Galam model of opinion dynamics and bottom-up hierarchical
voting. A different probability for a flip against the local majority
within the discussion group is associated with each ratio of majority
/ minority. The cases of groups of sizes 3 and 5 are investigated
in detail. For hierarchical voting, the local flip corresponds to
a `faithless elector', a representative who decides to vote against
the choice of their electing group. Depending on the flip probabilities,
the model exhibits a rich variety of patterns for the dynamics, which
include novel features in the topology of the landscape. In particular,
for size 5, we uncover for the first time an interplay between five
fixed points, which split into either three attractors and two tipping
points or two attractors and three tipping points, depending on the
flip probabilities. Larger groups are also analysed. These features
were absent in the former versions of the Galam model, which has at
maximum three fixed points for any group size. The results shed a
new light on a series of social phenomena triggered by one single
individual who acts against the local majority.
\end{abstract}


\section{Introduction}

The study of opinion dynamics has been and still is one of the main
topics of interest within the field of sociophysics \cite{ChChCh2006,CFL2009,Schw2019}.
Several recent surveys of the opinion dynamics literature are available
\cite{Red2019,Bra2020,Noo2020}. Models of opinion dynamics can be
classified according to the opinion space, which is the set of possible
opinions, and by whether time is discrete or continuous (the former
is more common; a recent example of the latter is found in \cite{SMA2020}).
In the 1980s and 90s, most models featured discrete opinion spaces.
The scenario where there are two choices has been especially thoroughly
studied (an influential article is \cite{Szn2000}). One of the main
contributions to this field is the Galam model \cite{Gal1986}, in
which individuals hold a certain binary opinion. Much later, continuous
opinion spaces were introduced such as the so called bounded confidence
models around the year 2000 (e.g.\! \cite{GNAW2000}), the Friedkin-Johnsen
model \cite{FJ1990,FJ1999}, and the Hegselmann-Krause model \cite{HK2002}.

Those continuous models connect to earlier modelling of opinion dynamics
by social psychologists and mathematicians starting in the 1950s \cite{As1956,Ab1964,DeGroot},
an approach which remained disconnected from sociophysics and physicists
for a few decades due to the compartmentalised nature of the different
disciplines and the lack of access to literature from other disciplines
pre internet. An increasing number of mathematicians have been joining
the field recently \cite{LT2013,AMBG2016,BBP2017}.

The basic Galam model describes the dynamics of binary opinions. We
call the options $A$ and $B$, with an initial proportion of $p_{0}\in\left[0,1\right]$
in favour of $A$. The model can be interpreted in two different ways:
\begin{enumerate}
\item As a model of \emph{voting in a hierarchy}. The hierarchy is a homogeneous
rooted tree (with branching factor $r\in\mathbb{N}$) with the leaves
being the bottom level individuals. Each group within the bottom level
is comprised of $r$ individuals who vote for one of the options by
the majority rule. This decision is carried to the next level of the
hierarchy, where $r$ representatives form a group that in turn votes
on $A$ versus $B$. Their decision is passed to the next level, etc.
With this setup, the number of voters on the bottom level of an $n$-hierarchy
with group size $r$ is $r^{n}$ and the total number of individuals
comprising the hierarchy is $N:=\left(r^{n+1}-1\right)/\left(r-1\right)$.
This model was first introduced in \cite{Gal1986}. See also \cite{Gal2017}
for an in-depth treatment.
\item As a model of \emph{opinion dynamics}. Rather than voting within fixed
groups as part of a hierarchy, the entire population randomly meets
in groups of fixed size $r$. A discussion takes place in each group
and the majority opinion is adopted by all members of the group. Next,
the groups are broken up and the individuals randomly form new groups
of size $r$ to resume discussion. This is repeated $n$ times. Contrary
to the model of voting in a hierarchy, the total number of individuals
is not a function of the number of discussion rounds $n$. However,
when studying the dynamics of the model, the number of rounds $n$
plays the same role as the number of levels in the hierarchy in the
voting model to determine the updated distribution of opinions. As
a rule of thumb, the total number of individuals $N$ should be at
a minimum 100 when dealing with proportions of opinions, since this
allows us to round to two digits. The influence of contrarians on
opinion dynamics was first studied using this model in \cite{Gal2002,Gal2004}.
There is also a version of `global' contrarianism \cite{BG2006},
which takes into account the global rather than the local (or group)
majority. Other authors have also studied the phenomenon of contrarianism,
e.g.\! \cite{TM2013,BR2015,GC2017,LHL2017,Ham2018}. The idea of
using contrarianism to explain social phenomena is not exclusive to
the study of opinion dynamics, however: \cite{CEMMS2002} is a study
of contrarianism in finance.
\end{enumerate}
In both cases, the interest lies in determining the dynamics of distribution
of preferences as $n$ increases. Aside from the basic model, there
have been several extensions \cite{Gal2002,Gal2017,GCh2020,GCh2020b}.
The authors of \cite{MR2003} studied a model with group size 3 and
probabilistic adoption of the group majority. In this paper, we compare
the dynamics of the basic opinion dynamics model, the contrarian model,
and a new local flip model, where the likelihood of an individual
flipping against the majority is a function of the magnitude of this
majority. We will assume exclusively that the size of the group $r\in\mathbb{N}$
is odd. This is motivated by the observation that if we introduce
a tiebreaker in case of a draw for even $r$, where each option is
then chosen with probability $1/2,$ then the dynamics of the model
correspond to that of the model with $r-1$ odd. See equation (2.8)
in \cite{Gal2017}.

A number of problems have been studied over the course of the last
decades using the general setup of the Galam model with suitable modifications.
Versions of this model have been used to predict surprising election
results, such as the French rejection of the European constitution
in 2005 \cite{Lehir}, the Brexit vote \cite{Gal2018}, and Trump's
election for president in 2016 \cite{Gal2017b}. However, the prediction
of a second Trump victory in 2020 failed by a short margin \cite{Gal2021}.
From a mathematical point of view, these scenarios can be modelled
by the repeated application of an update equation that takes as input
the current distribution of opinions, given by the proportion of individuals
with opinion $A$, and outputs the new proportion of individuals with
opinion $A$. Thus, this is a dynamic model where, for any initial
distribution of opinions, we have a trajectory of distributions over
time. The properties of the update equation determine whether the
dynamics are convergent and, if so, where they converge to. This is
how the model explains the emergence of stable majorities in favour
of one of the options, or to the contrary, under what circumstances
the dynamics tend to an even split between the two options.

In this article, we introduce a new model based on the Galam model,
called the local flip model. It is in a similar vein as the contrarian
model first introduced in \cite{Gal2004}, which deals with opinion
dynamics under the presence of some individuals who reject the majority
opinion. However, while the contrarian model uses the same probability
for all local configurations for a shift against the majority including
the case of unanimity, here we assume a different probability for
each configuration as a function of the ratio of majority to minority.
We investigate the local flip model for two settings, which are:
\begin{enumerate}
\item In the hierarchical voting scenario (the `vertical frame'), the
group representative faces incentives to deviate from the majority
opinion for personal gain. Suppose there is some mechanism in place
to detect deviations from the majority, e.g.\! a statistical sampling
of opinions across all groups, which would allow detection of deviations
some of the time, and make detection more likely when the majority
is large. Then it is reasonable to assume that flip probabilities
are decreasing in the magnitude of the majority. The local flip corresponds
to a `faithless elector', a representative who decides to vote against
the choice of their electing group.
\item The other scenario is the opinion dynamics interpretation. In this
`horizontal frame', there are contrarian tendencies which hamper
the adoption of the local majority opinion. Here, there is no reason
to assume that these tendencies are stronger for smaller majorities.
In fact, contrarianism may be stronger when facing a larger majority.
Hence, different parameter ranges may be suited to the vertical and
to the horizontal frame.
\end{enumerate}
We study in detail the cases of groups of sizes 3 and 5 as a function
of the various flip probabilities. The associated dynamics exhibit
a rich variety of patterns including novel features which were absent
in the previous works. In particular, contrary to the Galam model
which yields at maximum three fixed points (two attractors and one
tipping point) for any size $r$, the flip model is found to exhibit
five fixed points for $r=5$. Moreover, modifying the flip probabilities
produces an unexpected interplay between the stabilities of the five
fixed points with a transformation of a set of three attractors and
two tipping points into a set of two attractors and three tipping
points. These new features persist for larger group sizes. The previous
models referred to as the \textquoteleft basic model\textquoteright{}
and the \textquoteleft contrarian model\textquoteright{} in this article
are recovered as sub-cases of the local flip model. The results shed
a new light on a series of social phenomena, which could not be explained
by a flat probability of contrarian behaviour. Counterintuitive strategies
can be designed to optimise outcomes for competing groups.

This paper is organised in six sections and an appendix. In Sections
\ref{sec:Basic-Model} and \ref{sec:Contrarian}, we provide a review
of the basic Galam model and the contrarian model, respectively. Thereafter,
in Section \ref{sec:Flip-Voting-Model}, we introduce the local flip
model. We first study the model when the groups are very small, specifically
group size 3, and then we turn to the larger group size 5. We consider
versions of the model where flips only occur when the majority has
some specific magnitude, as well as versions where there are flips
for different majority sizes with distinct probabilities. Section
\ref{sec:Comparison} discusses the three models, their similarities
and differences. The novelties yielded by the local flip model are
reviewed. Section \ref{sec:Conclusion} concludes the paper. Finally,
the Appendix contains some technical details as well as the case of
arbitrarily large group sizes.

\section{\label{sec:Basic-Model}Basic Model}

Mathematically speaking, all the models are completely determined
by their update equation. This equation describes what the distribution
of opinions is for the next level / round given a current distribution
$p$. Given a current probability $p\in\left[0,1\right]$ of a preference
for $A$, the updated probability of a preference for $A$ is given
by the polynomial
\begin{equation}
P_{r}\left(p\right):=\sum_{i=\frac{r+1}{2}}^{r}\left(\begin{array}{c}
r\\
i
\end{array}\right)p^{i}\left(1-p\right)^{r-i}.\label{eq:basic}
\end{equation}
As mentioned in the Introduction, we are interested in the dynamics
of the probability that a preference for option $A$ exists: $p_{0}\mapsto p_{1}=P\left(p_{0}\right)\mapsto\cdots$.
The key to understanding these dynamics is the analysis of the fixed
points of the update function $P_{r}$. It is well known that for
the basic model, the three fixed points are $0,1/2,1$ for any group
size $r$. The points $0$ and $1$ are attractors. The fixed point
$1/2$ is unstable even for $r=3$ and it becomes `more unstable'
the larger $r$ is. This can be seen by calculating the derivative
\[
P_{r}'\left(1/2\right)=\left(\frac{1}{2}\right)^{r-1}\sum_{i=\frac{r+1}{2}}^{r}\left(\begin{array}{c}
r\\
i
\end{array}\right)\left(2i-r\right).
\]
This derivative is strictly larger than $1$ and it is asymptotically
equivalent to $\sqrt{2r/\pi}$ as $r\rightarrow\infty$. Thus, for
any initial distribution of preferences for $A$ given by the probability
$p_{0}\neq1/2$, we have a tendency to move towards one of the two
attractors: If $p_{0}<1/2$, then $p_{n}\searrow0$, and if $p_{0}>1/2$,
then $p_{n}\nearrow1$ as $n\rightarrow\infty$. This means that the
convergence to the stable fixed points becomes faster the larger the
group size $r$ becomes. In the limit $r\rightarrow N$ where there
is a single large group, the convergence becomes immediate ($n=1$).

Generally, the number of iterations $n$ required to converge to the
fixed points is quite small as long as $p_{0}$ is not too close to
the repeller $1/2$. We present numerical convergence data, i.e.\!
 the orbits $p_{0},p_{1}=P\left(p_{0}\right),p_{2}=P\left(P\left(p_{0}\right)\right),\ldots$
rounded to two decimal digits for initial probabilities $p_{0}=0.4$
and $p_{0}=0.48$ in Table \ref{tab:Dynamics_basic_04} and \ref{tab:Dynamics_basic_048}.
The rounding to two digits has a probabilistic interpretation: If
$k$ is the smallest value such that $p_{k}=0$, i.e.\!  $p_{k}=0.00$,
it means that either $p_{k}=0.001,0.002,0.003,0.004$, which in turn
implies that for $n=k$, on average, out of 1000 trials, we will find
that $A$ wins -- instead of $B$ as expected -- 1, 2, 3, or 4 times,
respectively. We observe that -- even for small group sizes -- the
convergence toward the fixed point 0 is very fast. This is due to
the superstability ($P_{r}'\left(0\right)=P_{r}'\left(1\right)=0$)
of the fixed points $0,1$. So even though formally we only have convergence
as $n\rightarrow\infty$, in reality, small values of $n$ suffice
to approach the fixed point.

\begin{table}
\begin{centering}
\begin{tabular}{|c|c|c|c|c|c|c|c|}
\hline 
$r$ & $p_{0}$ & $p_{1}$ & $p_{2}$ & $p_{3}$ & $p_{4}$ & $p_{5}$ & $p_{6}$\tabularnewline
\hline 
\hline 
$3$ & 0.4 & 0.35 & 0.28 & 0.2 & 0.1 & 0.03 & 0\tabularnewline
\hline 
$5$ & 0.4 & 0.32 & 0.19 & 0.05 & 0 & 0 & 0\tabularnewline
\hline 
$7$ & 0.4 & 0.29 & 0.11 & 0 & 0 & 0 & 0\tabularnewline
\hline 
$9$ & 0.4 & 0.27 & 0.06 & 0 & 0 & 0 & 0\tabularnewline
\hline 
\end{tabular}
\par\end{centering}
\caption{\label{tab:Dynamics_basic_04}Dynamics of the Basic Model $(p_{0}=0.4)$}
\end{table}
\begin{table}
\begin{centering}
\begin{tabular}{|c|c|c|c|c|c|c|c|c|c|c|c|}
\hline 
$r$ & $p_{0}$ & $p_{1}$ & $p_{2}$ & $p_{3}$ & $p_{4}$ & $p_{5}$ & $p_{7}$ & $p_{8}$ & $p_{9}$ & $p_{10}$ & $p_{11}$\tabularnewline
\hline 
\hline 
$3$ & 0.48 & 0.47 & 0.46 & 0.43 & 0.4 & 0.35 & 0.28 & 0.2 & 0.1 & 0.03 & 0\tabularnewline
\hline 
$5$ & 0.48 & 0.46 & 0.43 & 0.37 & 0.27 & 0.12 & 0.02 & 0 & 0 & 0 & 0\tabularnewline
\hline 
$7$ & 0.48 & 0.46 & 0.41 & 0.3 & 0.13 & 0.01 & 0 & 0 & 0 & 0 & 0\tabularnewline
\hline 
$9$ & 0.48 & 0.45 & 0.38 & 0.23 & 0.03 & 0 & 0 & 0 & 0 & 0 & 0\tabularnewline
\hline 
\end{tabular}
\par\end{centering}
\caption{\label{tab:Dynamics_basic_048}Dynamics of the Basic Model $(p_{0}=0.48)$}
\end{table}

\section{\label{sec:Contrarian}Contrarian Model}

This model is more adapted to the study of opinion dynamics. As in
the basic model above, the population meets repeatedly in randomly
formed groups of fixed size $r$. Discussion takes place and all $r$
members adopt the majority opinion. However, there are some individuals
in the population who have a tendency to adopt the opposite opinion.
This happens independently of their own initial opinion with a probability
of $a$. The behaviour of the model is summed up for $r=3$ in Table
\ref{tab:Contrarian}.
\begin{table}

\centering{}\renewcommand{\arraystretch}{1.8}%
\begin{tabular}{|c|c|c|c|c|}
\hline 
Pre discussion & Probability & Post discussion & Post contrarian & Probability\tabularnewline
\hline 
\hline 
\multirow{2}{*}{$AAA$} & \multirow{2}{*}{$p^{3}$} & \multirow{4}{*}{$AAA$} & $AAA$ & $\left(1-a\right)^{3}$\tabularnewline
\cline{4-5} \cline{5-5} 
 &  &  & $AAB\,\cdot3$ & $3a\left(1-a\right)^{2}$\tabularnewline
\cline{1-2} \cline{2-2} \cline{4-5} \cline{5-5} 
\multirow{2}{*}{$AAB\,\cdot3$} & \multirow{2}{*}{$3p^{2}\left(1-p\right)$} &  & $ABB\,\cdot3$ & $3a^{2}\left(1-a\right)$\tabularnewline
\cline{4-5} \cline{5-5} 
 &  &  & $BBB$ & $a^{3}$\tabularnewline
\hline 
\multirow{2}{*}{$ABB\,\cdot3$} & \multirow{2}{*}{$3p\left(1-p\right)^{2}$} & \multirow{4}{*}{$BBB$} & $AAA$ & $a^{3}$\tabularnewline
\cline{4-5} \cline{5-5} 
 &  &  & $AAB\,\cdot3$ & $3a^{2}\left(1-a\right)$\tabularnewline
\cline{1-2} \cline{2-2} \cline{4-5} \cline{5-5} 
\multirow{2}{*}{$BBB$} & \multirow{2}{*}{$\left(1-p\right)^{3}$} &  & $ABB\,\cdot3$ & $3a\left(1-a\right)^{2}$\tabularnewline
\cline{4-5} \cline{5-5} 
 &  &  & $BBB$ & $\left(1-a\right)^{3}$\tabularnewline
\hline 
\end{tabular}\caption{\label{tab:Contrarian}Contrarian Model, $r=3$}
\end{table}
For arbitrary $r$, the update equation of the contrarian model is
given by
\begin{equation}
Q_{r,a}\left(p\right):=\left(1-2a\right)\sum_{i=\frac{r+1}{2}}^{r}\left(\begin{array}{c}
r\\
i
\end{array}\right)p^{i}\left(1-p\right)^{r-i}+a.\label{eq:contrarian}
\end{equation}

Depending on the contrarian probability $a$, the dynamics differ
from those of the basic model. While $1/2$ is still a fixed point,
as in the basic model, $0$ and $1$ are no longer fixed points. We
consider group size $r=3$ first. Instead of $0$ and $1$, for $a\in\left(0,1/6\right)$,
there are two stable fixed points $p_{\pm}$: $p_{-}\in\left(0,1/2\right)$
and, symmetrically, $p_{+}=1-p_{-}\in\left(1/2,1\right)$. Specifically,
these fixed points are located at
\begin{equation}
p_{\pm}=\frac{1-2a\pm\sqrt{12a^{2}-8a+1}}{2\left(1-2a\right)}.\label{eq:contrarian_p_pm}
\end{equation}
For higher values of $a\in\left[1/6,1/2\right)$, $1/2$ is the unique
fixed point and it is stable. Qualitatively similar results hold for
$r\geq5$. The critical value of the contrarian probability $a_{c}$,
at which the phase transition occurs, equals $1/6$ for $r=3$. $a_{c}$
is increasing in $r$ and $a_{c}\nearrow1/2$ as $r\rightarrow\infty$.
For each group size, the pattern we observe is a repeller at $1/2$
and two stable fixed points $p_{\pm}$ if $a<a_{c}$ and the probability
of contrarianism is low. For probabilities of contrarianism in the
interval $\left[a_{c},1/2\right)$, we have a unique stable fixed
point at $1/2$. For $a=1/2$, the convergence to the attractor $1/2$
is immediate. If the contrarian probability is even higher, we obtain
an interesting alternating pattern. For $a\in\left(1/2,1-a_{c}\right]$,
$1/2$ is still an attractor, just as for $a\in\left[a_{c},1/2\right)$.
For $a>1-a_{c}$, there is an alternating pattern of converging subsequences
towards $p_{-}$ and $p_{+}=1-p_{-}$, respectively.

Convergence toward the stable fixed points is fast. We illustrate
this in Table \ref{tab:Dynamics_contrarian} which is representative
of the convergence behaviour of the model in a wide range of parameter
values. The values of $p_{n}$ given are rounded to two digits once
again. For the chosen parameter value $a=3/5$, we are in the alternating
regime with $1/2$ being an attractor. As we see, even starting at
$p_{0}=0.02$ very far away from the attractor $1/2$, convergence
is very fast.
\begin{table}
\begin{centering}
\begin{tabular}{|c|c|c|c|c|c|c|c|}
\hline 
$r$ & $p_{0}$ & $p_{1}$ & $p_{2}$ & $p_{3}$ & $p_{4}$ & $p_{5}$ & $p_{6}$\tabularnewline
\hline 
\hline 
$3$ & 0.02 & 0.6  & 0.47  & 0.51 & 0.5 & 0.5 & 0.5\tabularnewline
\hline 
$5$ & 0.02 & 0.6 & 0.46  & 0.51 & 0.49 & 0.5 & 0.5\tabularnewline
\hline 
$7$ & 0.02 & 0.6  & 0.46 & 0.52 & 0.49 & 0.5 & 0.5\tabularnewline
\hline 
$9$ & 0.02 & 0.6 & 0.45 & 0.52 & 0.49 & 0.51 & 0.5\tabularnewline
\hline 
\end{tabular}
\par\end{centering}
\caption{\label{tab:Dynamics_contrarian}Dynamics of the Contrarian Model $(a=3/5,p_{0}=0.02)$}
\end{table}

\section{\label{sec:Flip-Voting-Model}Local Flip Model}

Now we turn to the new model. One of the problems we want to study
is the behaviour of voting in hierarchies also referred to as `bottom-up
voting'. We imagine a hierarchy with group size $r$. Each group
votes and the majority decision is passed up the hierarchy. However,
the group's representative on the next higher level may not respect
the vote. We will consider a scenario where the deviation probability
depends on the size of the majority. If $\left(r+1\right)/2$ voters
within the group vote for $A$ and $\left(r-1\right)/2$ vote for
$B$, and we count each vote for $A$ as a $+1$ and each vote for
$B$ as a $-1$, then there is a $+1$ voting margin. We will write
$S$ for the voting margin. Similarly, an $\left(r-1\right)/2$-$\left(r+1\right)/2$
majority in favour of $B$ corresponds to a voting margin of $S=-1$.
Therefore, a minimal majority corresponds to an absolute voting margin
$\left|S\right|$ of $1$. Similarly, any number of votes $i=0,1,\ldots,r$
in favour of $A$ is equivalent to a voting margin of $S=i-\left(r-i\right)=2i-r$.

Before we define and analyse the local flip model in full generality,
we take a look at small groups of size 3 and, subsequently, of size
5. These models can be analysed exhaustively with rigorous results
concerning almost all of their properties. In the Appendix, we will
discuss the general model with any group size $r$.

\subsection{\label{subsec:r=00003D3}Group Size $r=3$}

For simplicity's sake, we first consider the case of groups of size
$r=3$. Then there are only two tiers of voting margins: $\pm1,\pm3$,
and hence the model has only two parameters which we will call $a$
and $b$. Let $a$ be the probability of deviating from the majority
when the absolute voting margin equals $1$ and $b$ the corresponding
probability if $\left|S\right|=3$. The behaviour of the model is
summarised in Table \ref{tab:Flip-Voting-r=00003D3}.
\begin{table}
\centering{}\renewcommand{\arraystretch}{1.8}%
\begin{tabular}{|c|c|c|}
\hline 
Configuration & Group vote & Probability\tabularnewline
\hline 
\hline 
$AAA$ & $A$ & $\left(1-b\right)p^{3}$\tabularnewline
\hline 
 & $B$ & $bp^{3}$\tabularnewline
\hline 
$AAB\,\cdot3$ & $A$ & $\left(1-a\right)\cdot3p^{2}\left(1-p\right)$\tabularnewline
\hline 
 & $B$ & $a\cdot3p^{2}\left(1-p\right)$\tabularnewline
\hline 
$ABB\,\cdot3$ & $A$ & $a\cdot3p\left(1-p\right)^{2}$\tabularnewline
\hline 
 & $B$ & $\left(1-a\right)\cdot3p\left(1-p\right)^{2}$\tabularnewline
\hline 
$BBB$ & $A$ & $b\left(1-p\right)^{3}$\tabularnewline
\hline 
 & $B$ & $\left(1-b\right)\left(1-p\right)^{3}$\tabularnewline
\hline 
\end{tabular}\caption{\label{tab:Flip-Voting-r=00003D3}Local Flip Model, $r=3$}
\end{table}
The update equation of the model is
\[
R_{3,a,b}\left(p\right)=\left(-2+6a-2b\right)p^{3}+\left(3-9a+3b\right)p^{2}+\left(3a-3b\right)p+b.
\]

We mention here that the local flip model can also be understood to
represent opinion dynamics / hierarchical voting under exogenous shocks:
Imagine a group decision process where the group decides according
to the majority rule, and then there is an exogenous shock that potentially
modifies the decision and which is stochastically independent of the
individual opinions. Provided that the shock probabilities are symmetric
and only depend on the size of the majority but not its sign, i.e.\!
 which of the option $A$ or $B$ is favoured, we obtain precisely
the local flip model. The symmetry of the flip probabilities implies
that the model is `neutral' in the sense that there is no distinction
between the two options $A$ and $B$. If we flipped all initial opinions
$A\longleftrightarrow B$, then the winning opinion would flip as
well.

Due to the small group size, we can completely solve the model and
analyse its dynamics. If the flip probability $b$ is non-zero, then
$0$ and $1$ are not fixed points of the dynamics. Due to the symmetry
of the model, however, $1/2$ is always a fixed point. To determine
the regimes of the model, we partition the parameter space $\left[0,1\right]^{2}=\left\{ \left(a,b\right)\vert\;a,b\in\left[0,1\right]\right\} $
first according to the stability of the fixed point $1/2$ and then
according to the number of fixed points the model has.

As far as the stability of $1/2$ is concerned, we distinguish four
regions:
\begin{enumerate}
\item The unstable region $\mathbf{L}$ with monotonic dynamics, given by
the inequality $b<1/3-a$.
\item The stable region $\mathbf{M}_{1}$ with monotonic dynamics, given
by the inequalities $1/3-a<b<1-a$.
\item The stable region $\mathbf{M}_{2}$ with alternating dynamics, given
by the inequalities $1-a<b<5/3-a$.
\item The unstable region $\mathbf{H}$ with alternating dynamics which
lies above the line $b=5/3-a$.
\end{enumerate}
We can also determine what happens at the boundaries between these
regions. See the Appendix for the details.

\begin{figure}
\centering{}\includegraphics{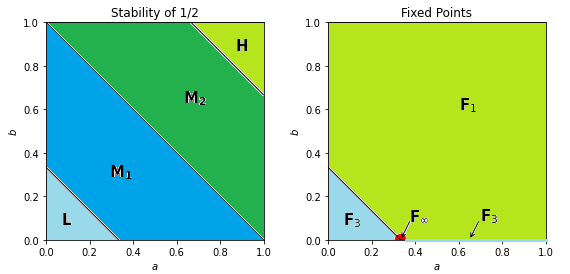}\caption{\label{fig:r=00003D3_regimes}Regimes of the Local Flip Model, $r=3$}
\end{figure}

Now we partition the parameter space into regions which exhibit either
a single fixed point or three fixed points, as well as a single point
where every value $p\in\left[0,1\right]$ is a fixed point.
\begin{enumerate}
\item The region in which every value $p\in\left[0,1\right]$ is a fixed
point is $\mathbf{F}_{\infty}:=\left\{ \left(1/3,0\right)\right\} $.
\item The region $\mathbf{F}_{1}$ with only a single fixed point which
is $1/2$ is given by the inequalities $1/3-a\leq b$ and $b>0$.
\item The region with three different fixed points is the complement $\mathbf{F}_{3}:=\left[0,1\right]^{2}\backslash\left(\mathbf{F}_{\infty}\cup\mathbf{F}_{1}\right)$,
i.e.\!  the corner region around the origin and the $a$-axis excluding
$\left(1/3,0\right)$.
\end{enumerate}
The regimes of the model can now be determined by forming intersections
of the regions given above.
\begin{enumerate}
\item The region $\mathbf{F}_{\infty}$ is a regime of its own. Here the
dynamics of the model are stationary.
\item The region $\mathbf{L}\subset\mathbf{F}_{3}$ is a regime characterised
by having three fixed points. $1/2$ is unstable and there are two
additional fixed points located at 
\[
p_{\pm}:=\frac{1}{2}\pm\frac{1}{2}\sqrt{\frac{1-3a-3b}{1-3a+b}}.
\]
Note that in the fraction both the numerator and the denominator are
positive. The points $p_{\pm}$ are attractors and their values are
not $0,1$ if and only if $b>0$. The location not being at $0,1$
is a new feature when compared to $b=0$ and similar to the contrarian
model where $a=b>0$.
\item The region $\mathbf{M}_{1}\cap\mathbf{F}_{1}$ exhibits a single fixed
point $1/2$ that is a global attractor. The dynamics of the model
starting at any $p_{0}$ is monotonic convergence towards $1/2$.
\item The region $\mathbf{M}_{1}\cap\mathbf{F}_{3}$, which is the section
of the $a$-axis with $a>1/3$, has three fixed points, $0,1/2,1$,
with $1/2$ being an attractor and $0,1$ repellers.
\item The region $\mathbf{M}_{2}\subset\mathbf{F}_{1}$ also has a single
fixed point $1/2$ that is a global attractor. However, the dynamics
of the model are dampened oscillations: starting at any $p_{0}$,
the orbits alternate around the limit $1/2$.
\item The region $\mathbf{H}\subset\mathbf{F}_{1}$ is characterised by
having a single fixed point $1/2$ which acts as a repeller. The dynamics
are oscillatory, so there are wild swings from one round to the next
from large majorities for $A$ to large majorities of $B$ and back.
In the extreme case $\left(a,b\right)=\left(1,1\right)$, we obtain
a model that is the inverse of the basic model: $R_{3,1,1}\left(p\right)=1-P_{3}\left(p\right)$.
The subsequences $p_{2n}$ and $p_{2n-1}$ converge and their limits
can be calculated explicitly:
\[
\omega_{\pm}:=\frac{1}{2}\pm\frac{1}{2}\sqrt{\frac{-10+6a+6b}{-2+6a-2b}}.
\]
These accumulation points are created at the centre, $p=1/2$, when
we cross over from $\mathbf{M}_{2}$ to $\mathbf{H}$ and they wander
towards the corners $0,1$ as the parameter values $\left(a,b\right)$
go toward the corner $\left(1,1\right)$. There is a period-doubling
bifurcation when passing from $\mathbf{M}_{2}$ to $\mathbf{H}$.
\end{enumerate}
A graph of the regimes can be found in Figure \ref{fig:r=00003D3_regimes}.
In the Introduction, we mentioned two scenarios: the vertical frame
is the study of hierarchical voting, and we consider that a representative
who decides whether to deviate from the group majority has a stronger
incentive to do so when the majority is of size 1, than when it is
of size 3. Thus, in this scenario, a reasonable assumption would be
$a>b$. In the horizontal frame, the group discussions may lead to
the phenomenon of contrarianism, where we consider that a larger majority
may strengthen the resolve of the contrarians. This interpretation
of the model leads to the assumption $a<b$. As we see in Figure \ref{fig:r=00003D3_regimes},
these two assumptions neatly bisect the two dimensional parameter
space, with the vertical frame being the lower right triangle and
the horizontal frame, the upper left triangle. Thus, with the exceptions
of $\mathbf{F}_{\infty}$ and $\mathbf{M}_{1}\cap\mathbf{F}_{3}$,
all other regions, both concerning the stability of 1/2 and the number
of fixed points overlap with both assumptions.

\subsection{Group Size $r=5$}

We will analyse various aspects of the model when the groups are of
size $r=5$. This model has three flip parameters: in addition to
$a$ and $b$ which are flip probabilities for $\left|S\right|$ equals
1 and 3, respectively, we have the flip parameter $c$ for $\left|S\right|=5$.
First, we will talk about a linear regime of the model which allows
us to determine the model's behaviour completely. Then, we will take
a look at the three one-parameter models where only one of $a,b,c$
is different than 0. We will subsequently compare these models to
the contrarian model which belongs to the category of one-parameter
models, too. To conclude this section, we will analyse the two-parameter
model with $c=0$ similarly to the discussion in Section \ref{subsec:r=00003D3}.

For groups of size 5, the update equation is
\begin{align}
R_{5,a,b,c}\left(p\right) & =\left(6-20a+10b-2c\right)p^{5}+\left(-15+50a-25b+5c\right)p^{4}\nonumber \\
 & \quad+\left(10-40a+30b-10c\right)p^{3}+\left(10a-20b+10c\right)p^{2}\nonumber \\
 & \quad+\left(5b-5c\right)p+c.\label{eq:R_5}
\end{align}
By varying all three flip parameters simultaneously, we obtain a `linear'
regime, i.e.\!  for certain values of the flip parameters, the update
function $R_{5,a,b,c}$ becomes an affine function of the form
\begin{equation}
R_{5,a,b,c}=\alpha+\beta p.\label{eq:aff_R}
\end{equation}
For $r=5$, we can determine the region in the parameter space $\left[0,1\right]^{3}$
explicitly for which $R_{5,a,b,c}$ has the form (\ref{eq:aff_R}).
From the update equation (\ref{eq:R_5}), we obtain the coefficients
of each power $p^{k},k=2,\ldots,5$, and equate them to 0. Thus, we
obtain a linear equation system, which we solve to obtain the solutions
given by
\begin{align*}
5a-c & =2\quad\textup{and}\quad5b-3c=1.
\end{align*}
This reduced equation system describes a line through parameter space,
along which the update function has the form (\ref{eq:aff_R}), with
the coefficients $\alpha=c$ and $\beta=1-2c$. We can parametrise
this line through parameter space as $\gamma:\left[0,1\right]\rightarrow\left[0,1\right]^{3}$
as a function of $c$. The first coordinate of $\gamma$ is $\gamma_{1}\left(c\right)=2/5+c/5=a$,
the second $\gamma_{2}\left(c\right)=1/5+3c/5=b$, and the last coordinate
is the value $\gamma_{3}\left(c\right)=c$. One way of looking at
this is to say that for each possible value of flip parameter $c$,
we have a unique configuration of all parameters that yields the linear
regime. Some values of $\gamma$ are given in Table \ref{tab:lin_reg}.
We will call this region of the parameter space $\mathbf{Lin}=\gamma\left(\left[0,1\right]\right)$.
\begin{table}
\begin{centering}
\begin{tabular}{|c|c|c|c|c|c|c|c|c|c|c|c|}
\hline 
$c$ & $0$ & $0.1$ & $0.2$ & $0.3$ & $0.4$ & $0.5$ & $0.6$ & $0.7$ & $0.8$ & $0.9$ & $1$\tabularnewline
\hline 
\hline 
$a$ & 0.4 & 0.42 & 0.44 & 0.46 & 0.48 & 0.5 & 0.52 & 0.54 & 0.56 & 0.58 & 0.6\tabularnewline
\hline 
$b$ & 0.2 & 0.26 & 0.32 & 0.38 & 0.44 & 0.5 & 0.56 & 0.62 & 0.68 & 0.74 & 0.8\tabularnewline
\hline 
\end{tabular}
\par\end{centering}
\caption{\label{tab:lin_reg}Parameter Values of the Linear Regime $(r=5)$}
\end{table}
Except for the two extremes, $\left(a,b,c\right)=\left(0.4,0.2,0\right)$
and $\left(a,b,c\right)=\left(0.6,0.8,1\right)$, there is a unique
fixed point $1/2$ which is a global attractor. For $c\leq1/2$, the
dynamics are monotonic, whereas for $c>1/2$, the dynamics are alternating.

\begin{figure}
\centering{}\includegraphics[scale=0.7]{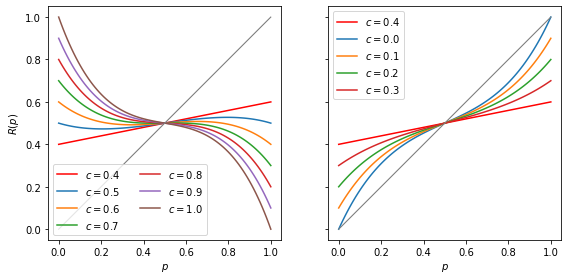}\caption{\label{fig:lin_regime}Linear Regime Separates Concave and Convex
$R_{5,a_{0},b_{0},c}$ with $a_{0}=0.48,b_{0}=0.44,c_{0}=0.4$}
\end{figure}

The existence of this linear regime allows us to find a result concerning
the fixed points of the update function $R_{5,a,b,c}$ given in (\ref{eq:R_5}):
fix a point $\left(a_{0},b_{0},c_{0}\right)\in\mathbf{Lin}$. Then
for all $c>c_{0}$, the update function $R_{5,a_{0},b_{0},c}$ is
strictly convex on the interval $\left[0,1/2\right]$ and strictly
concave on $\left[1/2,1\right]$. Similarly, if $c<c_{0}$, then $R_{5,a_{0},b_{0},c}$
is strictly concave on the interval $\left[0,1/2\right]$ and strictly
convex on $\left[1/2,1\right]$. In both cases, there are no fixed
points aside from $1/2$ and no 2-cycles (with the single exception
of the 2-cycle $\left(0,1\right)$ if $c=1$). Therefore, if $\left(a_{0},b_{0},c_{0}\right)\in\mathbf{Lin}$,
then, for all $c\in\left(0,1\right]$, $R_{5,a_{0},b_{0},c}$ has
no fixed points aside from $1/2$. See Figure \ref{fig:lin_regime}
for an illustration of this behaviour. It should be noted that --
generally speaking -- when all three parameters are distinct, $R_{5,a,b,c}$
will be neither concave nor convex over the intervals $\left[0,1/2\right]$
and $\left[1/2,1\right]$. 

Rigorous results concerning the existence and number of fixed points
aside from $1/2$ in this style are generally difficult to come by
for $r\geq5$ when all parameters assume distinct (non-zero) values.

In the right graph in Figure \ref{fig:lin_regime}, any value $0<c<0.4$
would serve as a model for a scenario that involves an initially dominant
opinion, and then, after one person deviates, others follow suit,
eventually leading to an even split.

Next, we consider models in which flips can only happen in the case
of a certain absolute voting margin.

\subsubsection{\label{subsec:b=00003Dc=00003D0}$\left|S\right|=1$}

Flips are now only possible for minimal majorities, i.e.\!  when
$\left|S\right|=1$, and $b=c=0$. The update equation (\ref{eq:R_5})
thus becomes
\begin{align}
R_{5,a}\left(p\right) & =\left(6-20a\right)p^{5}+\left(-15+50a\right)p^{4}+\left(10-40a\right)p^{3}+10ap^{2}\label{eq:R_5_b=00003Dc=00003D0}
\end{align}

The fixed points vary in number and stability depending on $a$. Here
is an overview:
\begin{enumerate}
\item For small $a\in\left[0,7/10\right]$, the model behaves similarly
to $a=0$, i.e.\!  the basic model. So $0,1/2,1$ are the only fixed
points and they have their usual stability properties.
\item If $a=3/10$, the update equation reduces to $R_{5,3/10}\left(p\right)=R_{3,0}\left(p\right)$.
\item At $a=7/10$, there is a phase transition.
\item For $a\in\left(7/10,1\right]$, there are five fixed points. The new
pair of fixed points is created at $1/2$, it is symmetric around
$1/2$, and located at
\begin{equation}
p_{\pm}:=\frac{1}{2}\pm\frac{1}{2}\sqrt{\frac{10a-7}{10a-3}}.\label{eq:p_pm_b=00003Dc=00003D0}
\end{equation}
Thus, we have a subcritical pitchfork bifurcation at $a=7/10$.
\item The fixed points $0$ and $1$ are always stable -- independently
of $a$. The fixed point $1/2$ is unstable in the regime $a\in\left[0,7/10\right]$
and stable for $a\in\left(7/10,1\right]$. The additional fixed points
$p_{\pm}$ are unstable.
\end{enumerate}
The behaviour with five fixed points only surfaces when there is a
large probability of flipping the vote.

This version of the model fits into the vertical frame interpretation
mentioned in the Introduction. We next investigate how the model changes
if we only consider flips at an absolute voting margin of 3.

\subsubsection{$\left|S\right|=3$}

Now only flips at absolute voting margins equal to 3 are considered.
So the model has a single flip parameter $b$. The update equation
(\ref{eq:R_5}) under this assumption is
\begin{align}
R_{5,b}\left(p\right) & =\left(6+10b\right)p^{5}+\left(-15-25b\right)p^{3}+\left(10+30b\right)p^{3}-20bp^{2}+5bp.\label{eq:R_5_a=00003Dc=00003D0}
\end{align}
The values $0,1/2,1$ are fixed points of $R_{r,b}$ for all values
of $b$. However, the number of fixed points and their stability depends
on $b$. Here is an overview:
\begin{enumerate}
\item For small $b\in\left[0,1/5\right]$ and for large $b\in\left[7/15,1\right]$,
the model has the three fixed points $0,1/2,1$. On the interval $\left(1/5,7/15\right)$,
there are five fixed points. The two additional fixed points are
\[
p_{\pm}:=\frac{1}{2}\pm\frac{1}{2}\sqrt{\frac{7-15b}{3+5b}}.
\]
\item For $b\leq1/5$, 0 and 1 are stable and $1/2$ is unstable. For all
$b>1/5$, the fixed points $0,1$ are unstable. However, the stability
of $1/2$ still varies for $b>1/5$.
\item On the interval $\left(1/5,7/15\right)$, $1/2$ is unstable and $p_{\pm}$
are stable. For $b\geq7/15$, $1/2$ is stable.
\end{enumerate}
It should be mentioned that this version of the model with $a=c=0$
does not fit into either the vertical or the horizontal frame interpretation
mentioned in the Introduction, because the finite sequence $a,b,c$
is neither increasing nor decreasing here (unless we disregard the
possibility of flips for $\left|S\right|=5$, in which case it fits
into the horizontal frame).

\subsubsection{$\left|S\right|=5$}

Now we consider a model with the possibility of a flip in case of
absolute voting margins equal to 5. So the flip parameters $a$ and
$b$ are equal to 0. The update equation is
\begin{align}
R_{5,c}\left(p\right) & =\left(6-2c\right)p^{5}+\left(-15+5c\right)p^{4}+\left(10-10c\right)p^{3}+10cp^{2}-5cp+c.\label{eq:R_5_a=00003Db=00003D0}
\end{align}
It should be noted that for $r=5$ the parameter $c$ is the flip
parameter of unanimous majorities. We observe that $0,1$ are fixed
points if and only if $c=0$.

\begin{figure}
\centering{}\includegraphics[scale=0.7]{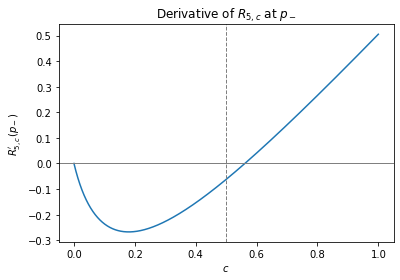}\caption{\label{fig:r=00003D5_derivative_p_-}Stability of $p_{\pm}$}
\end{figure}

Previously we saw that $a$ and $b$ could induce a phase transition.
The flip parameter $c$, on the other hand, cannot change the stability
of $1/2$ on its own. However, $c$ does change the location of the
other two fixed points. In fact, we observe that the fixed points
$p_{\pm}$, which lie between 0 and 1, wander towards $1/2$ as we
increase $c$. The range $\left[0,1\right]$ is not sufficient to
arrive at $1/2$. So, even for $c=1$, there are three fixed points.
It seems that $p_{\pm}$ are both stable throughout the entire range
of $c$ as we see in Figure \ref{fig:r=00003D5_derivative_p_-}. This
is related to the fact that $p_{\pm}$ are close to local extrema
of $R_{5,c}$, where the derivative is 0.

This version of the model fits into the horizontal frame interpretation
mentioned in the Introduction. 

\begin{figure}
\centering{}\includegraphics[scale=0.7]{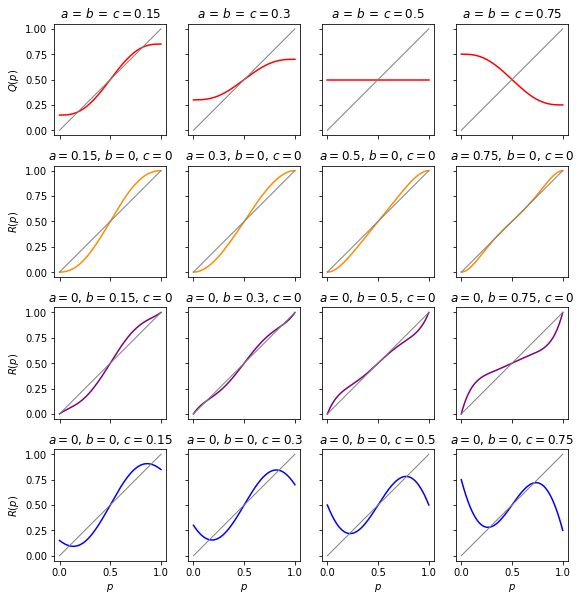}\caption{\label{fig:panel_one_param}Comparison of the Contrarian and One-Parameter
Local Flip Models}
\end{figure}

To conclude this part of the article, we compare the behaviour of
four one-parameter models: the contrarian model from Section \ref{sec:Contrarian}
and the $\left|S\right|=1,3,5$ models from this section. These graphs
are found in Figure \ref{fig:panel_one_param}. We see that both $a$
and $b$ can change the stability of the fixed point $1/2$. For small
values of these parameters, $1/2$ is unstable and $0,1/2,1$ are
the only fixed points. As the respective parameter increases, $1/2$
becomes stable and an additional pair of fixed points $p_{\pm}$ appears.
Interestingly, the mechanism by which the parameters create these
fixed points is different: the parameter $a$ leaves $0,1$ superstable
and only affects the stability of $1/2$. Thus, the unstable fixed
points $p_{\pm}$ appear at the point where $1/2$ switches to an
attractor. The parameter $b$ on the other hand affects the stability
of both $0,1$ and $1/2$. For small values of $b$, the fixed points
$0,1$ are still stable (but no longer superstable!) and $1/2$ is
unstable. At a critical value $b=1/5$, the fixed points $0,1$ become
unstable while $1/2$ is still unstable. This is only possible due
to the creation of two additional fixed points $p_{\pm}$ which are
stable. As $b$ further increases beyond $b=1/2$, the fixed point
$1/2$ becomes an attractor and $p_{\pm}$ disappear.

The parameter $c$, however, cannot on its own switch $1/2$ from
a repeller to an attractor. In the Appendix, we analyse the effect
of flip parameters for the general $r$ model. In equation (\ref{eq:lambda}),
we define the stability parameter $\lambda_{r}$ which is the derivative
of the update function $R_{r}$ at $1/2$. The finding that $c$ cannot
affect the stability of 1/2 is in line with the derivatives of $\lambda_{r}$
with respect to the flip parameter $a_{r}$ found in Table \ref{tab:lambda_r}.
(In the notation $a_{i}$, $i$ stands for the number of votes in
favour of $A$; thus, for $r=5$, $a=a_{3},b=a_{4},c=a_{5}$.) Note
that even for the case considered here where $r=5$, the partial derivative
$\partial\lambda_{5}/\partial c$ is $-0.62$. Since this partial
derivative does not depend on the value of $c$ according to (\ref{eq:partial_lambda_r}),
this explains why $c$ cannot affect the stability of $1/2$: even
going from $c=0$ to $c=1$ does not change $\lambda_{5}$ enough.

\begin{figure}
\centering{}\includegraphics[scale=0.7]{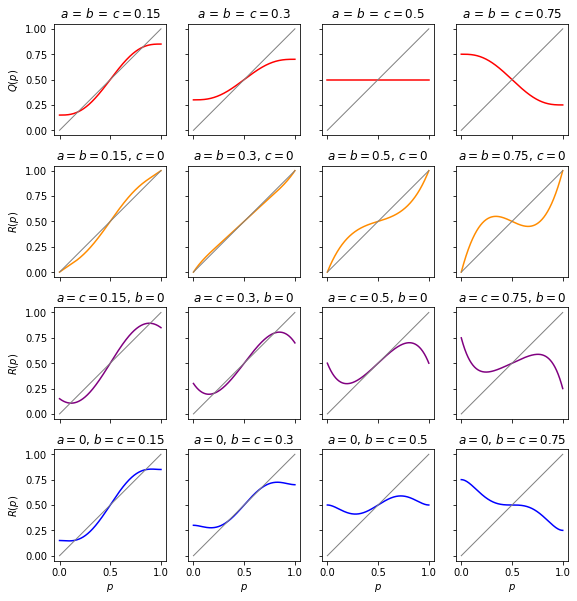}\caption{\label{fig:panel_two_param}Comparison of the Contrarian and Two-Parameter
Local Flip Models}
\end{figure}

\begin{figure}
\centering{}\includegraphics[scale=0.7]{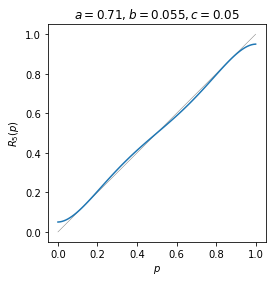}\caption{\label{fig:half-stable}Half-Stable Fixed Points in Local Flip Models}
\end{figure}

In Figure \ref{fig:panel_two_param}, we compare the contrarian model
to three different two-parameter models, in which we set one of the
three flip parameters $a,b,c$ equal to 0 and allow the other two
to vary. We note that the model with $a=b=0$ is qualitatively similar
for any value of $c$ to the contrarian model with low contrarian
probability.
In Figure \ref{fig:half-stable}, we present a scenario where all
three parameters are non-zero and the model is very close to $R_{5,a,b,c}\left(p\right)=p$
but there are three fixed points. The additional fixed points $p_{\pm}$
are half-stable: they are stable from the outside, i.e.\!  from the
left in case of $p_{-}=0.13$ and from the right for $p_{+}=0.87$,
and unstable from the inside, i.e.\!  from the direction of $1/2$.
This gives rise to a pattern where $p_{0}\in\left(0,p_{-}\right)$
implies convergence to $p_{-}$, $p_{0}\in\left(p_{-},p_{+}\right)$
leads to convergence to $1/2$, and $p_{0}\in\left(p_{+},1\right)$
towards $p_{+}$. The dynamics are monotonic in all cases. Both the
models with $b=0$ and $a=0$ can explain phenomena where initially
one of the opinions dominates completely, and then, after one person
deviates, others follow suit, with convergence either to some intermediate
point $p_{\pm}$ or 1/2.

\subsubsection{\label{subsec:small_majorities_r=00003D5}$\left|S\right|\protect\leq3$}

Now we analyse a two-parameter model that includes flips for absolute
voting margins equal to 1 and 3. That is $a$ and $b$ lie in $\left[0,1\right]$
and $c=0$. For this case, we can solve the model in similar depth
for $r=5$ as we did for $r=3$ in Section \ref{subsec:r=00003D3}.
The update equation is 
\begin{align}
R_{5,a,b,c}\left(p\right) & =\left(6-20a+10b\right)p^{5}+\left(-15+50a-25b\right)p^{3}\nonumber \\
 & \quad+\left(10-40a+30b\right)p^{3}+\left(10a-20b\right)p^{2}+5bp.\label{eq:R_5_c=00003D0}
\end{align}

Contrary to $r=3$, $0,1/2,1$ are always fixed points for $r\geq5$.
We proceed as in Section \ref{subsec:r=00003D3}, and partition the
parameter space $\left[0,1\right]^{2}$ first according to the stability
of the fixed point $1/2$ and then according to the number of fixed
points the model has.

As far as the stability of $1/2$ is concerned, we distinguish four
regions:
\begin{enumerate}
\item The unstable region $\mathbf{L}$ with monotonic dynamics, given by
the inequality $b<7/15-2a/3$.
\item The stable region $\mathbf{M}_{1}$ with monotonic dynamics, given
by the inequalities $7/15-2a/3<b<1-2a/3$.
\item The stable region $\mathbf{M}_{2}$ with alternating dynamics, given
by the inequalities $1-2a/3<b<23/15-2a/3$.
\item The unstable region $\mathbf{H}$ with alternating dynamics which
lies above $b=23/15-2a/3$.
\end{enumerate}
A similar remark as for $r=3$ regarding the inclusion of the boundaries
applies. The analysis is found in the Appendix.

\begin{figure}
\centering{}\includegraphics{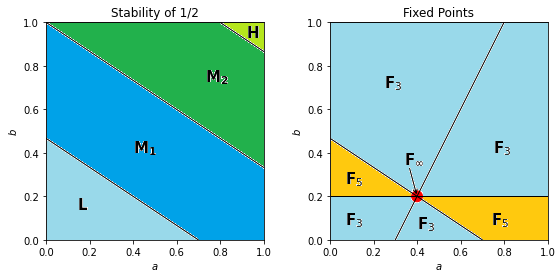}\caption{\label{fig:r=00003D5_regimes}Regimes of the Local Flip Model ($\left|S\right|\protect\leq3$),
$r=5$}
\end{figure}

Now we partition the parameter space into regions according to the
number of fixed points.
\begin{enumerate}
\item The region in which every value $p\in\left[0,1\right]$ is a fixed
point is $\mathbf{F}_{\infty}:=\left\{ \left(2/5,1/5\right)\right\} $.
\item The region $\mathbf{F}_{5}$ which has five fixed points consists
of the union of the two areas $1/5<b<7/15-2a/3$ and $7/15-2a/3<b<1/5$.
\item The region $\mathbf{F}_{3}$ with only the fixed points $0,1/2,1$
is the complement of $\mathbf{F}_{\infty}\cup\mathbf{F}_{5}$.
\end{enumerate}
The regimes of the model can now be determined by forming intersections
of the regions given above.
\begin{enumerate}
\item The region $\mathbf{F}_{\infty}$ is a regime of its own. Here the
dynamics of the model are stationary.
\item The region $\mathbf{L}\cap\mathbf{F}_{3}$ is a regime characterised
by having the three fixed points $0,1/2,1$. $1/2$ is unstable and
$0,1$ are stable. This is the regime most similar in behaviour to
the basic model since here the flip probabilities are low.
\item The region $\mathbf{L}\cap\mathbf{F}_{5}$ has five fixed points:
there are two additional fixed points located at 
\begin{equation}
p_{\pm}:=\frac{1}{2}\pm\frac{1}{2}\sqrt{\frac{7-10a-15b}{3-10a+5b}}.\label{eq:p_pm_c=00003D0}
\end{equation}
Note that in the fraction both the numerator and the denominator are
positive. As $1/2$ is unstable, the points $p_{\pm}$ are attractors
and $0,1$ are repellers. This is a feature the one-parameter model
with $b=c=0$ does not exhibit: if $b=0$, then the points $p_{\pm}$
are unstable.
\item The region $\mathbf{M}_{1}\cap\mathbf{F}_{3}$ has the fixed points
$0,1/2,1$. Here, $1/2$ is stable and the dynamics are monotonic.
\item The region $\mathbf{M}_{1}\cap\mathbf{F}_{5}$ has five fixed points,
$0,1/2,1$, and $p_{\pm}$ given by the formula above. $0,1/2,1$
act as attractors and $p_{\pm}$ as repellers.
\item The region $\mathbf{M}_{2}\subset\mathbf{F}_{3}$ also has the fixed
points $0,1/2,1$ with the same stability properties as $\mathbf{M}_{1}\cap\mathbf{F}_{3}$.
However, the dynamics of the model starting at any $p_{0}$ are now
alternating with limit $1/2$.
\item The region $\mathbf{H}\subset\mathbf{F}_{3}$ is characterised by
having the fixed points $0,1/2,1$ which acts as a repeller. The dynamics
are alternating, so there are wild swings from one round to the next
from large majorities for $A$ to large majorities of $B$ and back.
$0,1$ are fixed points contrary to the corresponding regime for $r=3$,
so if $p_{0}$ is even slightly different than $0$ or $1$, the dynamics
tend to produce large swings. There are accumulation points of the
dynamics in this regime provided $p_{0}\in\left(0,1\right)$. Similarly
to $r=3$, these are symmetric $\omega_{-}\in\left(0,1/2\right)$
and $\omega_{+}=1-\omega_{-}$. However, as $0,1$ are fixed points,
not even in the extreme case $\left(a,b\right)=\left(1,1\right)$
do we see the complete unanimity alternating between $A$ and $B$.
Instead, if the initial $p_{0}>0$ is very close to the origin, then
there is some lead time before the oscillation between majorities
for $A$ and $B$ starts. The discrepancy between the group sizes
$3$ and $5$ is because for the latter we are not allowing flips
for unanimous configurations.
\end{enumerate}
A graph of the regimes can be found in Figure \ref{fig:r=00003D5_regimes}.

The regime $\mathbf{L}\cap\mathbf{F}_{5}$ exhibits five fixed points:
two attractors $p_{\pm}$ given by (\ref{eq:p_pm_c=00003D0}) which
lie inside $\left(0,1/2\right)$ and $\left(1/2,1\right)$, respectively.
So the dynamics tend to majorities for one of the options. If only
$a$ is allowed to vary (as in Section \ref{subsec:b=00003Dc=00003D0}),
then there is no such regime: the only regime which exhibits five
fixed points has different stability properties. The additional fixed
points given by (\ref{eq:p_pm_b=00003Dc=00003D0}) act as tipping
points that push the dynamics of the model either to a fifty-fifty
split (if $p_{-}<p_{0}<p_{+}$) or towards unanimous opinions (if
$\left|p_{0}-1/2\right|>p_{+}-1/2$). Also, the regime $\mathbf{L}\cap\mathbf{F}_{5}$
appears at lower flip probabilities than the five fixed point regime
in the model with $b=c=0$.

In the Introduction, we mentioned two scenarios for the local flip
models: in the vertical frame, a reasonable assumption would be $a>b$.
The horizontal frame interpretation leads to the assumption $a<b$
(if we disregard the possibility of flips for $\left|S\right|=5$).
As we see in Figure \ref{fig:r=00003D5_regimes}, the situation is
similar to $r=3$, as these two assumptions bisect the two dimensional
parameter space, with the vertical frame being the lower right triangle
and the horizontal frame, the upper left triangle. Once again, with
the exception of $\mathbf{F}_{\infty}$, all other regions, both concerning
the stability of 1/2 and the number of fixed points overlap with both
assumptions. However, contrary to $r=3$, here we have two additional
fixed points, and as mentioned in the last paragraph, their stability
properties differ in the left hand side $\mathbf{L}\cap\mathbf{F}_{5}$,
where $p_{\pm}$ are attractors. This region belongs mostly to the
horizontal frame interpretation, although a small part of it lies
in the lower right triangle of the vertical frame. However, the right
hand side region $\mathbf{M}_{1}\cap\mathbf{F}_{5}$, where $p_{\pm}$
are repellers, belongs entirely to the vertical frame interpretation
of the model. Thus, contrary to $r=3$, the two interpretations of
the model, and their accompanying assumptions on the flip parameters,
do lead to different behaviours of the model. In particular, in the
two regimes with five fixed points each, the vertical frame induces
dynamics that tend to sizeable (but not unanimous) majorities in favour
of one of the two options. In the horizontal frame, the dynamics for
the most part either tend to an even split if $p_{0}$ starts between
$p_{\pm}$ and 1/2, or toward unanimous majorities if $p_{0}$ starts
between 0 and $p_{-}$ or between $p_{+}$ and 1.

\section{\label{sec:Comparison}Novel Behaviour of the Local Flip Model --
A Comparison of the Three Models}

We have defined three models in the previous three sections of this
article. In this section, we first take a look at the behaviour of
each of the three models, and then we describe the specific novel
aspects of the local flip model which do not occur in the other two
models. Each model is completely determined by its respective update
equation given in (\ref{eq:basic}), (\ref{eq:contrarian}), and (\ref{eq:flip_voting}),
and the model's properties can be deduced by analysing these update
equations.

\subsection{Basic Model}

We have seen that the basic model does not exhibit a phase transition.
For any update group size $r\in\IN$, the model exhibits three fixed
points. These are $0,1/2,1$. The stability properties of these fixed
points do not depend on $r$. Instead, 0 and 1 are always stable,
and 1/2 is always unstable. For any initial probability $p_{0}\neq1/2$
of a preference for $A$, the distribution of opinions tends to the
closer of the two attractors, either $0$ or $1$. The third fixed
point acts as a repeller. This behaviour is stable over all group
sizes $r$. In fact, as $r$ increases, the dynamics towards the attractors
become faster.

We can sum up the basic model by stating that there are stable unanimous
fixed points and a tipping point 1/2 which separates the two basins
of attraction.

For the other two models, we have observed that there are phase transitions
depending on the parameters, which in both models measure a probability
that some kind of deviation from the majority occurs.

\subsection{Contrarian Model}

In the contrarian model, each individual decides whether to go along
with the majority opinion or oppose it, regardless of their own initial
opinion. The contrarian model is a one-parameter model. It exhibits
a phase transition for all group sizes $r$ although the critical
value $a_{c}$ increases with the group size. In the low contrarian
probability regime, i.e.\!  $a<a_{c}$, the model behaves similarly
to the basic model with the main difference being the shift of the
stable fixed points from $0$ and $1$ to the points $p_{\pm}$ given
in equation (\ref{eq:contrarian_p_pm}), which lie in the interior
of the interval $\left[0,1\right]$. The change of $p_{\pm}$ is continuous
with respect to the parameter value $a$ on the interval $\left[0,a_{c}\right]$.
In the intermediate contrarian regime $a\in\left[a_{c},1-a_{c}\right]$,
the stable fixed points $p_{\pm}$ merge with the repeller $1/2$,
and this fixed point becomes globally stable. For high contrarian
probabilities, $a>1-a_{c}$, $1/2$ is once again a repeller, and
it is the only fixed point of the model. However, as opposed to the
low regime, $p_{n}$ does not converge to either $p_{-}$ or $p_{+}$;
instead, there is an alternating pattern with subsequences converging
to $p_{\pm}$. Similarly to the basic model, the contrarian model
does not exhibit more than three fixed points for any group size.
Also, we note that the contrarian model does not feature $p_{\pm}$
as unstable fixed points. As we saw, the local flip model does have
this distinction.

\subsection{Local Flip Model}

We considered two different scenarios to be explained by the local
flip model. One is hierarchical voting, where the representative,
whose task it is to pass along the majority decision of their group,
sometimes decides to flip to the contrary opinion but does so taking
into consideration the size of the majority. The other scenario is
opinion dynamics, where each person decides on their own whether to
adopt the majority opinion.

For group size $r$, the local flip model spans a $\left(r+1\right)/2$-dimensional
parameter space, in contrast to the basic and the contrarian model,
which are restricted to zero- and one-dimensional parameter spaces,
respectively. Exploring the multidimensional parameter space of the
local flip model reveals a wide variety of different and novel behaviours.
In particular, depending on the region in the parameter space, we
found dynamics which are driven by respectively one, three, and --
for the first time -- five fixed points for groups of size $r=5$.
The finding of five fixed points is new and significant in that, for
both the basic and the contrarian models, all sizes $r$ for the discussion
groups always yield only one or three valid fixed points. The others
roots of the polynomial of degree $r$ that is the update function
given by (\ref{eq:basic}) and (\ref{eq:contrarian}), respectively,
lie outside the interval $\left[0,1\right]$.

While $1/2$ is always a fixed point, its stability is contingent
on the parameter values. Moreover, in the local flip model, the fixed
points 0 and 1 can become unstable, contrary to what happens in the
other two models where they are always attractors. Likewise the additional
fixed points $p_{\pm}$, which lie respectively in the interior of
the intervals $\left(0,1/2\right)$ and $\left(1/2,1\right)$, can
be stable or unstable. These changes of stability create a series
of counterintuitive dynamics with flow toward or away from 1/2 and
$p_{\pm}$. The phenomena of convergence to large majorities, ties,
or alternating majorities are obtained within one single framework.
 Within the local flip model, both the basic and the contrarian model
appear as special cases. In addition, new dynamics are found, which
do not occur in those two models. For instance, unlike in the contrarian
model, where $p_{\pm}$ are always attractors, in the local flip model
$p_{\pm}$ can be tipping points directing the dynamics towards either
$1/2$ or one of the unanimous fixed points 0 and 1.

This new feature highlights a novel strategy for the competing alternatives
$A$ and $B$. For $A$ to avoid a loss, an initial support $p_{0}>p_{-}$
is required; otherwise, the dynamics tend to 0 and a unanimous majority
in favour of $B$. When $1/2>p_{0}>p_{-}$, the support for $A$ increases
to reach a tie at $1/2$. However, getting a high initial support
$p_{+}>p_{0}>1/2$ is futile and even a waste since then the dynamics
tend down to $1/2$. To trigger an increase in support, $A$ must
reach an initial support $p_{0}>p_{+}$. These results suggest a twofold
strategy to either reach a tie or to win, with an interval of support
$\left(1/2,p_{+}\right)$ which is illusionary in the sense that it
does not suffice to win in the end.

We summarise the properties of the three models in Table \ref{tab:Summary}.
\begin{table}
\begin{centering}
\renewcommand{\arraystretch}{1.5}%
\begin{tabular}{|c|c|c|c|c|c|c|}
\hline 
\multirow{2}{*}{} & \multirow{2}{*}{Basic} & \multirow{2}{*}{Contrarian} & Local flip & Local flip & Local flip & Local flip\tabularnewline
 &  &  & $b=c=0$ & $a=c=0$ & $a=b=0$ & $c=0$\tabularnewline
\hline 
\hline 
$0,1$ fixed points & $\checkmark$ & $\Circle$ & $\checkmark$ & $\checkmark$ & $\Circle$ & $\checkmark$\tabularnewline
\hline 
$0,1$ stable & $\checkmark$ & $\checkmark$\tablefootnote{\label{fn:0_1_fixed}Provided $0,1$ are fixed points.} & $\checkmark$ & $\Circle$ & $\checkmark^{\text{\ref{fn:0_1_fixed}}}$ & $\Circle$\tabularnewline
\hline 
$1/2$ fixed point & $\checkmark$ & $\checkmark$ & $\checkmark$ & $\checkmark$ & $\checkmark$ & $\checkmark$\tabularnewline
\hline 
$1/2$ unstable & $\checkmark$ & $\Circle$ & $\Circle$ & $\Circle$ & $\checkmark$ & $\Circle$\tabularnewline
\hline 
Phase transition & $\times$ & $\checkmark$ & $\checkmark$ & $\checkmark$ & $\times$ & $\checkmark$\tabularnewline
\hline 
Min $\#$ fixed points & 3 & 1 & 3 & 3 & 3 & 1\tabularnewline
\hline 
Max $\#$ fixed points & 3 & 3 & 5 & 5 & 3 & 5\tabularnewline
\hline 
Alternating pattern & $\times$ & $\Circle$ & $\times$ & $\times$ & $\times$ & $\Circle$\tabularnewline
\hline 
\end{tabular}\caption{\label{tab:Summary}Summary of the Models for $r=5$ ({\small{}$\checkmark$
`yes', $\Circle$ `sometimes', $\times$ `no')}}
\par\end{centering}
\end{table}

\section{\label{sec:Conclusion}Conclusion}

In this article, starting from the Galam model of opinion dynamics
which uses local majority rules to update individual opinions in small
discussion groups, we have included probabilistic deviations from
the local majority as a function of the ratio majority / minority,
which resulted in the local flip model of opinion dynamics and hierarchical
voting. Investigating groups of size 3 and 5 has revealed a series
of novel features which were absent in the previous works. In particular,
for group size 5, the dynamics are driven by five fixed points, which
split into three attractors and two tipping points or two attractors
and three tipping points, depending on the parameter values. Such
an unexpected and new interplay of fixed point stabilities creates
a rich diversity of non-linear dynamics, which could shed some new
light on several social phenomena triggered by one or a few individuals
acting against larger local majorities, such as:
\begin{itemize}
\item At a party, initially people are standing around, having conversations.
Then at some point in time, the first person starts dancing and others
follow. After a short time, some fixed proportion of all people in
attendance are dancing. This illustration was considered using a different
model \cite{SznW2021_priv,JMNSz-W2018}. Note that the iteration can
consist of the people moving around the room and thus reshuffling
the update groups in which conversations take place. Also, every new
song which is played can be considered a new round of discussions
as it changes the dynamics.
\item During a demonstration, one person starts throwing objects at the
police and others close by follow suit. Rioters move among demonstrators
and drag more and more people into violence. Also, in this scenario,
the movement in the crowd leads to iterations with different configurations
of small groups that influence each other.
\item Sharing some confidential or sensational piece of information, be
it political, societal, or financial in nature, can cause a few people
to shift their opinion. They will keep their new stance after the
meeting and subsequently even use the new information to convince
others in subsequent meetings. This includes the spreading of fake
news.
\end{itemize}

In future work, we intend to introduce asymmetric flip probabilities
along the same lines as the asymmetric contrarianism probabilities
in \cite{GCh2020}. Another direction we will consider is the case
of large size update groups with correlated initial opinions, which
could be more appropriate to describe some social situations involving
crowd behaviour.

\section*{Appendix}

\subsection*{Boundary Analysis between Regimes}

We analysed the two-parameter model for $r=3$ in Section \ref{subsec:r=00003D3}.
As mentioned there, we can determine the behaviour of the model on
the boundary between the stability regions of the universal fixed
point 1/2. On the boundary between the regions $\mathbf{L},\mathbf{M}_{1},\mathbf{M}_{2},$
and $\mathbf{H}$, the first derivative of the update function $R_{3,a,b}'\left(1/2\right)$
has some critical value: $-1,0,$ or 1. Therefore, it is not possible
to determine the stability just by looking at the first derivative.
The second derivative at $1/2$ is 0, so we go to the third derivative.
The line $b=-1+3a$ represents the location of all combinations of
$a$ and $b$ such that the third derivative $R_{3,a,b}'''\left(1/2\right)$
equals $0$. To the left of this line, the third derivative is negative;
to the right, it is positive. Thus, to the left of $b=-1+3a$, the
boundaries between $\mathbf{L}$ and $\mathbf{M}_{1}$, $\mathbf{M}_{1}$
and $\mathbf{M}_{2}$ belong to the higher one of the regions. To
the right of the line, the pattern is opposite and the boundaries
belong to the lower one of the neighbouring regions.

Similarly, for the two-parameter model with $r=5$ and $c=0$ analysed
in Section \ref{subsec:small_majorities_r=00003D5}, we can sort out
the behaviour on the boundaries. The line $b=1-2a$ represents the
location of all combinations of $a$ and $b$ such that the third
derivative $R_{5,a,b}'''\left(1/2\right)$ equals $0$. To the left
of this line, the boundary between $\mathbf{L}$ and $\mathbf{M}_{1}$
belongs to $\mathbf{M}_{1}$. To the right of the line, the pattern
is opposite and the boundaries belong to the lower one of the neighbouring
regions.

\subsection*{Larger Group Sizes}

We discuss the general model of group size $r$. This model has a
flip probability $a_{i}$ for any number of votes belonging to the
majority, i.e.\!  $i=\left(r+1\right)/2,\ldots,r$, and symmetrically
for a majority of the same respective magnitude for $B$. The update
equation is given by
\begin{equation}
R_{r,\mathbf{a}}\left(p\right)=\sum_{i=\frac{r+1}{2}}^{r}\left(\begin{array}{c}
r\\
i
\end{array}\right)\left[\left(1-a_{i}\right)p^{i}\left(1-p\right)^{r-i}+a_{i}p^{r-i}\left(1-p\right)^{i}\right].\label{eq:flip_voting}
\end{equation}
The parameters $a_{i}$ are the flip probabilities when there are
$i$ votes for $A$ (or, symmetrically, $i$ votes for $B$). Just
as in the case of the contrarian model, the local flip model also
reduces to the basic model if all parameters are 0: $\mathbf{a}=\left(a_{\left(r+1\right)/2},a_{\left(r+3\right)/2},\ldots,a_{r}\right)=\left(0,\ldots,0\right)$.
Due to the symmetry of the model, $1/2$ is a universal fixed point
for any group size. Furthermore, the unanimous points $0,1$ are fixed
if and only if $a_{r}=0$, i.e.\!  the probability that the group
representative deviates from a unanimous group vote is 0. What is
more, the values are
\begin{equation}
R_{r,\mathbf{a}}\left(0\right)=a_{r}\quad\textup{and}\quad R_{r,\mathbf{a}}\left(1\right)=1-a_{r}.\label{eq:R_0_1}
\end{equation}

Since disregarding a unanimous vote might easily lead to widespread
discontent, assuming $a_{r}=0$ may be reasonable. In fact, this is
called the Pareto criterion in the voting theory literature; see e.g.\!
Chapter 1 of \cite{TayPac}. The derivatives of $R_{r,\mathbf{a}}$
at $0,1$ are given by
\begin{equation}
R'_{r,\mathbf{a}}\left(0\right)=R'_{r,\mathbf{a}}\left(1\right)=r\left(a_{r-1}+a_{r}\right).\label{eq:lambda_0_1}
\end{equation}
Thus, only the two parameters of the unanimous and unanimous-but-one
flip affect the stability of $0,1$.

If $a_{r-1}=a_{r}=0$, then $0,1$ are superstable fixed points. Thus,
there is some neighbourhood of $0$ or $1$ in which convergence towards
$0$ or $1$ is very fast.

As we saw in Section \ref{subsec:r=00003D3}, even varying a single
flip parameter can switch the model into another regime when $r=3$.
Now we turn to the question of how much a single flip parameter can
influence the dynamics of the model when the group size is larger.
Specifically, we analyse the stability of $1/2$. We define
\begin{align}
\lambda_{r} & :=R'_{r,\mathbf{a}}\left(1/2\right)=\frac{1}{2^{r-1}}\sum_{i=\frac{r+1}{2}}^{r}\left(\begin{array}{c}
r\\
i
\end{array}\right)\left(1-2a_{i}\right)\left(2i-r\right).\label{eq:lambda}
\end{align}
We observe that $\lambda_{r}$ can be expressed as the sum of the
derivative of the update function of the basic model, $P'_{r}\left(1/2\right)$,
and another term
\begin{equation}
-\frac{1}{2^{r-2}}\sum_{i=\frac{r+1}{2}}^{r}\left(\begin{array}{c}
r\\
i
\end{array}\right)a_{i}\left(2i-r\right)\label{eq:lambda_summand_R}
\end{equation}
which is negative. The stability parameter $\lambda_{r}$ of the fixed
point $1/2$ can be regarded as a differentiable function of the flip
parameters $a_{i}$: $\lambda_{r}:\left[0,1\right]^{\frac{r+1}{2}}\rightarrow\mathbb{R}$
Thus, we will also write $\lambda_{r}\left(\mathbf{a}\right)$. The
partial derivatives
\begin{equation}
\frac{\partial\lambda_{r}}{\partial a_{i}}=-\frac{1}{2^{r-2}}\left(\begin{array}{c}
r\\
i
\end{array}\right)\left(2i-r\right)\label{eq:partial_lambda_r}
\end{equation}
are all negative, whereas $\lambda_{r}$ can be positive or negative.
This means that in regimes where $\lambda_{r}$ is positive, increasing
$a_{i}$ contributes to making $1/2$ more stable. In regimes with
alternating dynamics around $1/2$, increasing $a_{i}$ makes $1/2$
less stable. We present some of the values in Table \ref{tab:lambda_r}.
As we see, the flip parameter with the largest impact is not $a_{\left(r+1\right)/2}$
but rather some $a_{i}$ with $i$ close to $\left(r+1\right)/2$.
So, to change the dynamics, flips of small -- but not minimal --
majorities are most effective. We also observe that the flip parameter
$a_{r}$, which by (\ref{eq:R_0_1}) completely determines the values
$R_{r,\mathbf{a}}\left(0\right)=a_{r}\textup{ and }R_{r,\mathbf{a}}\left(1\right)=1-a_{r}$,
only marginally affect the stability of the universal fixed point
$1/2$. Lastly, note that $\partial\lambda_{r}/\partial a_{i}$ is
constant with respect to any of the flip parameters.
\begin{table}
\begin{centering}
\begin{tabular}{|c||c|c|c|c|c|c|c|}
\hline 
\multirow{2}{*}{$r$} & \multicolumn{7}{c|}{$\partial\lambda_{r}/\partial a_{i}$, where $i=$}\tabularnewline
 & $\left(r+1\right)/2$ & $\left(r+3\right)/2$ & $\left(r+5\right)/2$ & $\left(r+7\right)/2$ & $\left(r+9\right)/2$ & $\left(r+11\right)/2$ & $\left(r+13\right)/2$\tabularnewline
\hline 
\hline 
$3$ & $-1.5$ & $-1.5$ & -- & -- & -- & -- & --\tabularnewline
\hline 
$5$ & $-1.25$ & $-1.88$ & $-0.62$ & -- & -- & -- & --\tabularnewline
\hline 
$7$ & $-1.09$ & $-1.97$ & $-1.09$ & $-0.22$ & -- & -- & --\tabularnewline
\hline 
$9$ & $-0.98$ & $-1.97$ & $-1.41$ & $-0.49$ & $-0.07$ & -- & --\tabularnewline
\hline 
11 & $-0.9$ & $-1.93$ & $-1.61$ & $-0.75$ & $-0.19$ & $-0.02$ & --\tabularnewline
\hline 
13 & $-0.84$ & $-1.89$ & $-1.75$ & $-0.98$ & $-0.34$ & $-0.07$ & $-0.01$\tabularnewline
\hline 
15 & $-0.79$ & $-1.83$ & $-1.83$ & $-1.17$ & $-0.5$ & $-0.14$ & $-0.02$\tabularnewline
\hline 
17 & $-0.74$ & $-1.78$ & $-1.89$ & $-1.32$ & $-0.65$ & $-0.23$ & $-0.05$\tabularnewline
\hline 
19 & $-0.7$ & $-1.73$ & $-1.92$ & $-1.45$ & $-0.8$ & $-0.33$ & $-0.1$\tabularnewline
\hline 
\end{tabular}
\par\end{centering}
\caption{\label{tab:lambda_r}Partial Derivatives of $\lambda_{r}$ (Rounded
to Two Digits)}
\end{table}
In fact, we can characterise the behaviour of the index $i$ with
the most negative partial derivative. Asymptotically, i.e.\!  for
large $r$, the index $i$ with the most negative partial derivative
$\partial\lambda_{r}/\partial a_{i}$ is $i\approx r/2+\sqrt{r}/4$.
Here, we used the symbol `$\approx$' in the sense that two sequences
$f\left(n\right),g\left(n\right)$ are asymptotically equivalent,
i.e.\!  $f\left(n\right)\approx g\left(n\right)$, if $\lim_{n\rightarrow\infty}f\left(n\right)/g\left(n\right)=1$.

However, each flip parameter individually does not have a significant
effect on the stability of $1/2$ as $r$ grows large. As we will
see, for $r$ as small as 13, no single parameter on its own can turn
$1/2$ stable. Hence, the statement concerning the most impactful
flip parameter should be interpreted in the sense that a set of flip
parameters with indices close to $r/2+\sqrt{r}/4$ jointly have the
most impact.

The situation differs with respect to the fixed points $0,1$. We
already know from (\ref{eq:lambda_0_1}) that only the two parameters
$a_{r-1}$ and $a_{r}$ can affect the stability of these fixed points.
Since $a_{r}>0$ implies that $0,1$ are not fixed points in the first
place, we see that $a_{r-1}$ determines on its own whether $0,1$
are stable or unstable. Given $a_{r}=0$, the points $0,1$ are stable
if and only if $a_{r-1}\leq1/r$. Interestingly, as the group size
becomes larger, even small values of the flip parameter $a_{r-1}$
can make $0,1$ unstable and induce a tendency toward smaller majorities
when starting at some $p_{0}$ arbitrarily close to either 0 or 1.

How come a single parameter can affect the stability of $0,1$ but
the same is not possible for $1/2$? The slope parameter $\lambda_{r}$
at $1/2$ is positive for $\mathbf{a}=0$ which corresponds to the
basic model. Here, the term (\ref{eq:lambda_summand_R}) is 0. As
$r$ becomes larger, this slope $\lambda_{r}\left(0\right)$ becomes
larger as well, behaving like $\lambda_{r}\left(0\right)=P'_{r}\left(1/2\right)\approx\sqrt{2r/\pi}$
as noted in Section \ref{sec:Basic-Model}. So even though the largest
derivatives in absolute value found in Table \ref{tab:lambda_r} and
given by the formula (\ref{eq:partial_lambda_r}) do \emph{not} decay
as $r$ increases, their magnitude compared to $\lambda_{r}\left(0\right)$
becomes insignificant. The same does not apply to the fixed points
$0,1$: the slope of $R_{r,\mathbf{a}}$ at these points is 0 for
$\mathbf{a}=0$. That, in combination with the multiplicative constant
$r$ in (\ref{eq:lambda_0_1}), allows a single parameter to affect
the stability of these fixed points even (and especially) when $r$
is large.

What are the ranges of $r$ such that each of three parameters $a=a_{\left(r+1\right)/2},b=a_{\left(r+3\right)/2},c=a_{\left(r+5\right)/2}$
can induce a phase transition on their own?
\begin{itemize}
\item The flip parameter $a$ can induce a phase transition if and only
if $r\leq5$. For larger values of $r$, even $a=1$ does not make
the fixed point 1/2 stable.
\item The flip parameter $b$ is more effective at producing a phase transition
in the model: it can turn 1/2 stable for group sizes up to and including
$r=11$. The critical values are given in Table \ref{tab:b_c}. For
$r\geq13$, there is no phase transition. Here, for all values $b\in\left[0,1\right]$,
the dynamics of the model resemble those of the basic model, i.e.\!
 the case where $b=0$.
\item The flip parameter $c$ cannot induce a phase transition even for
$r=5$. (For $r=3$, there is no flip parameter $c$, only $a$ and
$b$.)
\end{itemize}
\begin{table}
\begin{centering}
\begin{tabular}{|c|c|c|c|c|}
\hline 
$r$ & $5$ & $7$ & $9$ & $11$\tabularnewline
\hline 
\hline 
$b_{c}\left(r\right)$ & 0.47 & 0.6 & 0.74 & 0.88\tabularnewline
\hline 
\end{tabular}
\par\end{centering}
\caption{\label{tab:b_c}Critical Values for $b$ (Rounded to two Digits)}
\end{table}

\end{document}